\def\section{\@startsection {section}{1}{\z@}{-26pt plus -1ex minus
    -.2ex}{13pt plus .2ex}{\large \bf}}
\def\subsection{\@startsection{subsection}{2}{\z@}{-13pt plus -1ex minus
   -.2ex}{13pt plus .2ex}{\large \rm}}
\def\subsubsection{\@startsection{subsubsection}{3}{\z@}{-13pt plus
  -1ex minus -.2ex}{-9pt plus .2ex}{\large \em}}
\def\paragraph{\@startsection
     {paragraph}{4}{\z@}{3.25ex plus 1ex minus .2ex}{-1em}{\large\bf}}
\def\subparagraph{\@startsection
     {subparagraph}{4}{\parindent}{3.25ex plus 1ex minus
     .2ex}{-1em}{\large\bf}}

\renewcommand{\author}[3]{\vspace{5mm}
\begin{center}
{\normalsize \rm #1}\\    
{\normalsize \it #2}\\    
{\normalsize \it #3}\\    
\vspace{0.35cm}\end{center} }
            
\def\lsim{~\rlap{$<$}{\lower 1.0ex\hbox{$\sim$}}}
\def\gsim{~\rlap{$>$}{\lower 1.0ex\hbox{$\sim$}}}

\def \aa#1#2   {{\em Astr. Astrophys. \/} {\bf #1}, {#2}}
\def \aas#1#2  {{\em Astr. Astrophys. Suppl. Ser. \/} {\bf #1}, {#2}}
\def \aj#1#2   {{\em Astron. J. \/} {\bf #1}, {#2}}
\def \apj#1#2  {{\em Astrophys. J. \/} {\bf #1}, {#2}}
\def \apjs#1#2 {{\em Astrophys. J. Suppl. Ser. \/} {\bf #1}, {#2}}

\def \nat#1#2  {{\em Nature \/} {\bf #1}, {#2}}

\documentclass[aps,superscriptaddress,altaffilletter,tightenlines]{revtex4}

\newcommand{\be}{\begin{equation}} 
\newcommand{\ee}{\end{equation}} 
\newcommand{\tres}[1]{\buildrel\ldots\over #1} 

\newcommand{\n}{\label}

\begin{document}

\title{ASYMPTOTIC BEHAVIOUR OF BIANCHI VI~$_0$ SOLUTIONS WITH AN EXPONENTIAL-POTENTIAL
SCALAR FIELD}


\author{Luis P. Chimento}\email{chimento@df.uba.ar}
\affiliation{ Departamento de F\'{\i}sica, Facultad de Ciencias Exactas y
Naturales,  Universidad de Buenos Aires, Ciudad
Universitaria, Pabell\'on I, 1428 Buenos Aires, Argentina}

\author{Pablo Labraga}\email{wtblalop@lg.ehu.es}
\affiliation{Departamento de F\'\i sica Te\'orica, Universidad del Pa\'\i s Vasco, Apartado 644, 
48080 Bilbao, Spain.}

\begin{abstract}

\vskip 0.5cm

\baselineskip .34cm
{\bf Abstract}\,\,\,\,\,We obtain some solutions to the Einstein-Klein-Gordon equations without a 
cosmological constant for an exponential potential scalar field in a
Bianchi~VI$_0$ metric and investigate their behaviour.
\end{abstract}

\maketitle

\vskip 0.5cm

\section{Introduction}
Inflationary theories are expected to explain the observed isotropy of
the Universe by assuming an inflationary expansion in the early
universe, since as the so-called ``Cosmic no-hair" theorems state, all
the solutions of the Einstein equations initially expanding and with a
positive cosmological constant evolve towards the de Sitter solution
\cite{wald}, \cite{jensen}. However, only those Bianchi models that have the FRW models 
as particular solutions isotropize~\cite{heusler}. One of the
potentials which has received special attention is that of Liouville
form (exponential potential), the one we will use in this work.

Some anisotropic cosmological models have already been studied. 
Aguirregabiria, Feinstein and Ib\'a\~nez~\cite{afi} analyzed the
Bianchi~I models with exponential potential by reducing the problem of
finding exact solutions to the solution of one third order differential
equation. In \cite{ach} a non-local transformation was used to linearize
this equation and the general solution was found. In this work we shall
extend this analysis to the Bianchi~VI$_0$, reducing, again, the problem
of finding exact solutions to resolve a nonlinear differential equation
and investigating the behaviour of their solutions at early time and
its asymptotic stability in the far future. 

The line-element for a Bianchi~VI$_0$ cosmological model can be
written in the following form:
\be 
\n{1}
ds^2=e^{f(t)}\left( -dt^2+dz^2\right) + G(t)\left( e^zdx^2+e^{-z}dy^2\right) . 
\ee 
and the corresponding Klein-Gordon and Einstein field equations for the
metric~(\ref{1}) are as follows:
\be
\n{3}
\ddot\phi +\frac{\dot G}{G}\dot\phi + e^f\frac{\partial V}{\partial\phi}=0,
\ee 
\be 
\n{4}
\frac{\ddot G}{G}=2e^fV, 
\ee 
\be 
\n{5}
\frac{\ddot G}{G}-\frac{1}{2}\frac{\dot G^2}{G^2}-\dot f\frac{\dot G}{G}+ 
\frac{1}{2}+\dot\phi^2=0. 
\ee 

We will consider a homogeneous self-interacting scalar field
($\phi=\phi(t)$) with an exponential potential $V=\Lambda e^{k\phi}$,
which is the only potential that separates the Klein--Gordon equation in
two parts, one containing only the scalar field and the other containing
only geometrical quantities (see \cite{paper}). Making use of this potential 
we can express the scalar field as follows 
\be
\n{9}
\dot\phi=\frac{m}{G}-\frac{k}{2}\frac{\dot G}{G}. 
\ee 
After some algebra, the above equations can be reduced to the following
one:
\be 
\n{general}
\ddot G^2 G-K\ddot G\dot G^2-\tres G\dot G G+\frac{1}{2}\ddot G G^2 + m^2\ddot G=0;
\ee 
where $K=k^2/4-1/2$. This is the equation we are going to study in detail. First we
will consider the behaviour of the solutions when $t\rightarrow 0$ and then we will 
study their stability.

\section{Behaviour of the solutions when $t\rightarrow 0$}

\subsection{Particular case}

If we consider Equation~(\ref{general}) when $m=0$, and we assume that
the function $G$ has the form $G(t)=G_0 (\Delta t)^n$ when $\Delta
t\rightarrow 0$ ($\Delta t = t-t_0$), we find two types of solutions,
one with $n=1$, which leads to a vanishing potential, and the other with
$n=1/K$. In the special case of $-\frac{1}{2}<K<0$ we obtain a power law
inflationary solution (see \cite{lm}), which can be written in its 
synchronous form as follows:
\be 
\n{infla}
ds^2 = - dT^2 + \beta T^{\frac{2 n}{n+2}} \left[ e^{Z/\sqrt{\alpha}} dX^2 +
e^{-Z/\sqrt{\alpha}} dY^2 + dZ^2 \right] , 
\ee 
where $\beta = {[(n+2)/(2\sqrt{\alpha})]}^{2/(n+2)}$, $\alpha =\frac{n(n-1)}
{2\Lambda |\phi_0|}$ and $\phi_0$ a constant.

We have studied the asymptotical behaviour of the scalar curvature,
whose expression for the metric (\ref{1}) is:
\be
\n{R}
R=e^{-f(t)}\left[ \frac{1}{2}+\frac{1}{2}\frac{\dot{G}^2}{G^2}-
2\frac{\ddot{G}}{G}-\ddot{f}\right].
\ee

In this particular case, we obtain that $R$ diverges at $t=0$ for $K>0$,
so we can speak about a singularity at that point. In the case of
negative $K$ the scalar curvature vanishes when $t\rightarrow 0$, due to
the fact that $t\rightarrow 0$ means $T\rightarrow \infty$ in
Equation~(\ref{infla}). 

\subsection{General Case}

In the most general case the equation we have to deal with is:
\be 
\label{enes} 
\frac{1}{(\Delta t)^4}\left[n^2-(1+K)n^3+Kn^4\right]-\left(\frac{1}{2}+
\frac{m^2}{(\Delta t)^{2n}}\right)\frac{1}{(\Delta t)^2}\left[ n^2-
n\right]=0.
\ee 

When $n\le 1$ we obtain the same results that we had in the particular
case. Solutions with $n>1$, however, are not permitted. This means that 
solutions with $0>K>1$ do not appear in the general case.

As we did for the general case, we can calculate the behaviour of the
scalar curvature, given by equation~(\ref{R}). The relevant solutions
(those with $-\frac{1}{2}<K<0$ or $1\le K$) behave as they did in the
$m=0$ case, and therefore, the conclusions about the divergence of $R$
remain unchanged. 

\section{Stability of the asymptotic solution}

\subsection{Particular Case}

To investigate the behaviour and stability of the solutions, we can
rewrite Equation~(\ref{general}) in a quite different way, by
introducing a new function $h=\frac{\dot G}{G}$ and redefining the time
variable as follows:
\be
d\eta = hdt = d\ln{G}.
\ee
After some transformations Equation~(\ref{general}) reads
\be 
\n{infinito}
\frac{d}{d\eta}\left[\frac{h'^2}{2}+K\frac{h^2}{2}-\frac{1}{2}\ln{h}\right]=
-\left[-\frac{1}{2h^2}+(1+K)\right] h'^2,
\ee 
where primes are derivatives with respect to the new variable $\eta$.
This is the equation of motion for a dissipative or antidissipative
system, with the potential ${\cal V}(h)=K\frac{h^2}{2}-\frac{1}{2}\ln{h}$. 

Equation~(\ref{infinito}) presents local minima when $h_0^2=\frac{1}{2K}$, 
for $K>0$ (i.e. $k^2>2$). As the dissipative term given by the right-hand side 
of Equation~(\ref{infinito}) is negative definite in the asymptotic regime, the bracket on 
the l.h.s. of Equation~(\ref{infinito}) define a Liapunov Function \cite{cesari},
\cite{luis}, \cite{ale}. Then, the corresponding exact solution, expressed by 
\be 
\n{estable}
G=G_0e^{\sqrt\frac{1}{2K}\,t}
\ee 
(with $G_0$ a constant) is stable for $t\rightarrow \infty$ and for any 
initial condition. This result allows us to study the behaviour of the solutions 
around these equilibrium points. To first order in perturbations and for 
$K>-\frac{1}{2}$, there is a two-parameter family
of stable solutions that behaves as Equation~(\ref{estable}). The
trajectories in the phase plane $(h,\dot h)$ can be divided in two
different groups: for $K>\frac{1}{8}$ the solutions cut the axis in the phase plane an
infinite number of times, so they spiral around the constant solution
$h_0$ (Equation~(\ref{estable})). For $-\frac{1}{2}<K\le\frac{1}{8}$ the
solutions do not cut the $h$ axis or they cut it once.  

\subsection{General Case}

Equation~(\ref{general}) when $m=0$ can be written using $h$ and $\eta$
variables, as follows:
\be 
\label{infigen} 
\frac{d}{d\eta}\left[ \frac{h'^2}{2}+{\cal V}(h)\right] = 
2m^2e^{-2\eta}\ln{h}-\left\{(1+K)-\frac{1+2m^2e^{-
2\eta}}{2h^2}\right\}h'^2,
\ee 
where now the ``potential" is ${\cal V}(h)=\frac{Kh^2}{2}-\left\{ 
\frac{1}{2}+m^2e^{-2\eta}\right\} \ln{h}$, 
and the local minima will be given by $h_0^2=\frac{1+2m^2e^{-2\eta}}{2K}$. 
One has to be more careful in this case because the equilibrium points
are stable only if $K>\frac{1}{2}$, as a consequence of Liapunov's
theorem. In this case, the equilibrium points are not fixed, they depend
on $\eta$, and the equation that rules their behaviour gives us the
following solution:
\be 
G_{min}=\pm\left\{-\frac{m^2}{2}e^{\sqrt{\frac{1}{2K}}\,(t-t_0)} +
e^{-\sqrt{\frac{1}{2K}}\,(t-t_0)}\right\} ,
\ee 
so the final asymptotically stable solution behaves again as
Equation~(\ref{estable}), which represents an anisotropic solution. This
result had been obtained numerically in \cite{afi2} and, in our case, it
corresponds to the simplest solution we can obtain from
Equation~(\ref{general}) when $m=0$.

\section{Conclusions}

We have studied the general behaviour of Bianchi~VI$_0$ solutions with a
self-interacting scalar field and an exponential potential, by reducing
the system of field equations to one differential equation of third
order, Equation~(\ref{general}).

We have analyzed the behaviour of these solutions when $t\rightarrow 0$
and have found several anisotropic solutions which appear to have a
singularity at $t=0$, since the scalar curvature diverges there. A
power-law inflationary solution corresponding to $k^2<2$ has also been
found.

We have studied the asymptotical behaviour of the solutions when
$t\rightarrow\infty$, finding that there is a simple asymptotically
stable anisotropic solution for all cases. It can be easily shown
(see \cite{paper}) that this solution remains anisotropic for large $t$.

\acknowledgments
We wish to thank Prof. J.~Ib\'a\~nez and A.~Feinstein
for helpful discussions about this subject. P.L.'s work was supported by
the Basque Government fellowship B.F.I. 92/090. This work was supported
by the Spanish Ministry of Education grant (CICYT) No PB93-0507.

\end{document}